\documentclass[10pt, a4paper]{article}
\pdfoutput=1

\usepackage[utf8x]{inputenc}
\usepackage[LGR, T1]{fontenc}
\PrerenderUnicode{ä}
\PrerenderUnicode{é}

\usepackage{amsmath, amsthm, amsfonts, amssymb}
\usepackage{textcomp}
\usepackage{textgreek}
\usepackage{upgreek}
\usepackage{mathrsfs}

\usepackage[tt=false, type1=true]{libertine}
\usepackage[varqu]{zi4}
\usepackage[libertine]{newtxmath}
\usepackage{microtype}
\usepackage{geometry}

\usepackage[english]{babel}

\usepackage{xspace}

\usepackage{xcolor}
\definecolor{keywordcolor}{rgb}{0.7, 0.1, 0.1}   
\definecolor{commentcolor}{rgb}{0.4, 0.4, 0.4}   
\definecolor{symbolcolor}{rgb}{0.0, 0.1, 0.6}    
\definecolor{sortcolor}{rgb}{0.1, 0.5, 0.1}      

\usepackage{listings}

\usepackage{hyperref}
\hypersetup{
    colorlinks,
    linkcolor={red!50!black},
    citecolor={blue!50!black},
    urlcolor={blue!80!black}
   }
\usepackage[capitalize]{cleveref}

\newtheorem{theorem}{Theorem}[section]
\newtheorem{proposition}[theorem]{Proposition}
\newtheorem{lemma}[theorem]{Lemma}

\newcommand{\mathlib}{\texttt{mathlib}\xspace}

\newcommand{\nhds}[1]{\mathcal{N}_{#1}}

\long\def\U{\mathcal{U}}

\DeclareMathOperator{\Spa}{Spa}

\DeclareMathOperator{\Spv}{Spv}

\DeclareMathOperator{\Cont}{Cont}

\renewcommand{\O}[1][X]{\mathcal{O}_{#1}}

\newcommand{\F}{\mathscr{F}}

\long\def\G{\mathscr{G}}

\newcommand{\NN}{\mathbb{N}}

\newcommand{\ZZ}{\mathbb{Z}}

\newcommand{\QQ}{\mathbb{Q}}

\newcommand{\RR}{\mathbb{R}}

\newcommand{\separated}{separated\xspace}

\long\def\C{\mathcal{C}}

\DeclareMathOperator{\inv}{inv}

\DeclareMathOperator{\hatinv}{\widehat{\inv}}

\DeclareMathOperator{\supp}{supp}

\DeclareMathOperator{\Frac}{Frac}

\DeclareMathOperator{\pr}{pr}

\DeclareMathOperator{\res}{res}

\renewcommand{\subset}{\subseteq}

\title{Formalising perfectoid spaces}
\author{Kevin Buzzard \thanks{Imperial College London. Supported by EPSRC grant EP/L025485/1.}
 \and
 Johan Commelin \thanks{Universität Freiburg.
 Supported by the Deutsche Forschungs Gemeinschaft (DFG)
 under Graduiertenkolleg 1821 (Cohomological Methods in Geometry).}
 \and
 Patrick Massot
 \thanks{Laboratoire de Mathématiques d'Orsay,
         Univ. Paris-Sud, CNRS, Université Paris-Saclay}}

\begin{document}
\maketitle
\begin{abstract}
  Perfectoid spaces are sophisticated objects in arithmetic geometry
  introduced by Peter Scholze in 2012.
  We formalised enough definitions and theorems
  in topology, algebra and geometry
  to define perfectoid spaces in the Lean theorem prover.
  This experiment confirms that a proof assistant
  can handle complexity in that direction,
  which is rather different from formalising a long proof about simple objects.
  It also confirms that mathematicians
  with no computer science training can become proficient users of a proof assistant
  in a relatively short period of time.
  Finally, we observe that formalising a piece of mathematics
  that is a trending topic
  boosts the visibility of proof assistants amongst pure mathematicians.
\end{abstract}

\section{Introduction}

In 2012, Peter Scholze defined the notion of a perfectoid space,
and used it to prove new cases of the weight-monodromy conjecture,
an extremely important conjecture in modern arithmetic geometry.
This original application of the theory was based on
a key theorem of Scholze called the \emph{tilting correspondence},
relating perfectoid spaces in characteristic zero
to those in positive characteristic.
Over the next few years,
many other applications appeared,
culminating in some spectacular applications to the Langlands programme.
Scholze was awarded the Fields Medal in 2018 for his work.
See~\cite{rapoport, wedhorn_survey} for far more thorough explanations of
how Scholze's ideas have changed modern mathematics.

With current technology, it would take many person-decades to formalise
Scholze's results.
Indeed, even \emph{stating} Scholze's theorems would be an achievement.
Before that, one has of course to formalise the definition of a perfectoid space,
and this is what we have done,
using the Lean theorem prover.

For a quick preview, here is what the final definitions in our code look like.
\begin{lstlisting}
structure perfectoid_ring (A : Type) [Huber_ring A] extends Tate_ring A :=
(complete   : is_complete_hausdorff A)
(uniform     : is_uniform A)
(ramified   : ∃ ϖ : pseudo_uniformizer A, ϖ^p ∣ p in Aᵒ)
(Frobenius : surjective (Frob Aᵒ∕p))

def is_perfectoid (X : CLVRS) : Prop :=
∀x, ∃ (U : opens X) (A : Huber_pair) [perfectoid_ring A],
  (x ∈ U) ∧ (Spa A ≊ U)

def PerfectoidSpace := {X : CLVRS // is_perfectoid X}
\end{lstlisting}

Formalising the definitions above
(and in particular the meaning of $\Spa(A)$ and of the isomorphism
$\Spa(A)\cong U$)
requires formalising many intermediate definitions and proving many
theorems,
and was done without adding any axioms to Lean's logic.
Formalising the definition of such sophisticated objects seems rather different
from formalising a complex proof about simpler objects, as in \cite{odd_order},
but turns out to be also accessible to proof assistants.
Contemporary mathematics is also formalised in \cite{cap_set,
gouezel_morse}, but those papers focus on elementary objects (finite
fields, linear algebra and metric spaces).

The goal of this text is to report on the formalisation process,
highlighting our efforts to bridge the gap between the formal and informal
stories.
The level of details decreases quickly along the paper,
mostly because the involved mathematics becomes too technical for the
format constraints,
but also because the relatively elementary parts of the story probably
are interesting to more people.

As far as we know, no other formalisation of perfectoid spaces has been
attempted in any theorem prover.
A priori it could be done,
although it seems more difficult without dependent types.
For instance, sheaves are functions associating a ring to each open subset of a
topological space,
an idea which is naturally expressed as a dependent type.

Our main source for the classical part of the story is Bourbaki's
\emph{Elements of Mathematics}.
More recent sources all seem to focus on special cases that are not general
enough for our purposes,
and defer to Bourbaki for the general case.
And Bourbaki's careful style is also very valuable.
We hope our readers won't be misled by the fact that we do point out a couple
of arguments that we could improve on.
That's simply because repeating everywhere else that Bourbaki was spot on
does not make a very interesting story.
\looseness=-1

Our main source for the recent part of the story is Wedhorn's lecture notes
\cite{wedhorn},
with occasional need to go back to the original sources by Huber and Scholze
\cite{huber_valuations,huber_adic_spaces, scholze_thesis, scholze_diamonds}.

\section{Background and Overview}
\label{sec:background_and_overview}

A perfectoid space is an adic space satisfying some properties. The history of
adic spaces goes back to the basic question of how to do analysis over complete
fields such as the field $\QQ_p$ of $p$-adic numbers, and other fields of
interest to number theorists.
The classical theory of analysis of one complex variable has a local
definition of differentiation,
with powerful consequences such as the theorem that a differentiable function on
the complex plane has a power series expansion which converges everywhere.
Whilst there was clearly some $p$-adic analogue of the theory
(for example $p$-adic exponential and logarithm functions shared many
properties with their classical counterparts),
the foundations of the subject remained elusive for the simple reason that
$\QQ_p$ is totally disconnected,
and there are plenty of examples of differentiable functions which
were locally constant but not constant.
To give a specific example,
consider the function on the closed unit disc $\ZZ_p$
defined as the characteristic function of the open unit disc
(so~1 on the open disc and~$0$ elsewhere).
It is locally constant
(as both the open unit disc and its complement are open,
due to the non-Archimedean nature of the metric),
and hence satisfies the naive definition of differentiability,
however it has no global power series expansion.

In the late 1960s,
Tate proposed a foundation to the theory
which solved these problems,
namely the theory of rigid analytic spaces.
Let $\QQ_p\langle X\rangle$ denote
the ring of power series which converge on $\ZZ_p$.
Tate starts with the observation that each point $a$ in $\ZZ_p$
yields a maximal ideal of $\QQ_p\langle X\rangle$,
namely the kernel of the ring homomorphism $\QQ_p\langle X\rangle\to\QQ_p$
sending $f$ to $f(a)$. Tate's replacement for the closed unit
disc is the collection of all maximal ideals of $\QQ_p\langle X\rangle$.
Using ``spaces'' formed by maximal ideals of rings of convergent power series,
Tate developed a $p$-adic theory
analogous to the theory of real or complex manifolds.
This theory had all the right properties,
but the underlying ``spaces'' were no longer topological spaces,
they were spaces equipped with a Grothendieck topology.
Tate's resolution of the problem
with the characteristic function of the open unit disc
is that the cover of the rigid analytic closed unit disc
by the open disc and its complement is not
admissible for his Grothendieck topology,
even though both sets are open.

However, the use of a Grothendieck topology was inconvenient.
In the 1990s, Berkovich and Huber proposed variants of the theory.
The maximal ideals which made up Tate's ``spaces''
gave rise to valuations on the rings he introduced.
We can add more points to these spaces
by considering more general valuations on
these rings. Berkovich considered real-valued valuations, but Huber
allowed his valuations to take values in certain totally ordered monoids.
Then many extra points appear on or near the boundary of the open unit disc,
with the result that the function defined to be
$1$ on the open disc
and $0$ outside it
is no longer a continuous function;
the complement of the open disc is no longer open.
For a vivid description of the points that get added to the closed unit disc
over a complete algebraically closed field,
see \cite[Example 2.20]{scholze_thesis}.

Huber called his new objects \emph{adic spaces},
and the majority of our work is a formalisation of their definition.
Adic spaces represent a decisive generalisation of
Tate's theory of rigid analytic spaces,
because finiteness constraints
(such as finite-dimensionality of the objects in question)
are embedded in Tate's foundations,
but are not necessary in the theory of Berkovich or adic spaces.
This was crucial for Scholze,
whose perfectoid spaces are built from rings which almost never satisfy these
constraints.

Adic spaces have more structure than a topology.
They are endowed with a sheaf of complete topological rings,
and equivalence classes of valuations on the stalks of this sheaf.
In particular, this structure associates to each open subset $U$ in our space
a ring equipped with a topology
for which addition and multiplication are continuous,
and which is complete
in a sense that generalises the definition of complete metric spaces.
This ring is meant to describe analytic functions on $U$,
although its elements are not literally functions with domain~$U$.
Instead, the sheaf axioms enable us to manipulate these elements as if they
were functions.

In order to motivate the introduction of completeness in this story,
let us evoke how to go from a polynomial ring $k[X]$ to the corresponding
power series ring $k[[X]]$ from a topological point of view.
One can define a topology on $k[X]$, which is compatible with its ring structure,
and such that neighbourhoods of $0$ are subsets containing the ideal $X^nk[X]$
for some natural number $n$.
Then the partial sums $S_N$ of a power series can be seen as a sequence in $k[X]$
which is Cauchy for this topology:
the difference $S_N - S_M$ belongs to an arbitrarily small neighbourhood of zero
as $N$ and $M$ tend to infinity.
One can then view $k[[X]]$ as a completion of $k[X]$ where this sequence
converges.
This algebraic notion of completion,
present throughout our work,
corresponds to the geometric fact that we want
certain uniform limits of analytic functions on our spaces
to be analytic.

Adic spaces are constructed by gluing model spaces
(affinoid preadic spaces)
associated to certain kinds of topological rings.
This is analogous to how manifolds are locally modelled on open subsets of
affine spaces,
and schemes are locally modelled on spectra of rings.
But here the local models are more complicated to build,
and our construction of an object in a certain category {\tt CLVRS}
from a so-called \emph{Huber pair} of topological rings
involves proving a lot of theorems.
This is the bulk of our formalisation effort.
It involves a subtle blend of general topology and commutative algebra.

All the story so far brings us to the 90s.
Briefly, the motivation behind Scholze's perfectoid spaces is the following.
Some of the simplest examples of maps between geometric objects are the
unramified maps:
maps which are locally isomorphisms.
This geometric notion has an arithmetic counterpart;
the extensions of local fields which are easiest to study are the unramified
extensions.
Again it was Tate who showed us another approach;
if instead one considered so-called ``deeply ramified'' extensions,
then they were so badly ramified
everywhere that any finite extension of a deeply ramified extension was
``almost unramified''.
This idea was taken further first by Faltings,
and then by Scholze, who extended the theory to adic spaces:
Scholze's perfectoid spaces can be thought of as
the ``deeply ramified'' adic spaces. In particular, the {\tt Frobenius}
field of the {\tt perfectoid\_ring} structure above forces many elements
of the ring~$A$ to have $p$th roots, guaranteeing that in most
cases $A^\circ$ is non-Noetherian. This makes the algebra harder,
but what we gain is Scholze's tilting theorem, saying that in a precise
sense the geometry of a perfectoid space in characteristic zero
is controlled by its characteristic $p$ ``tilt'',
giving a profound link between geometry
in characteristic~$p$ and characteristic~0.
The details of this link are beyond the scope of this paper.
The uses of perfectoid spaces in mathematics are complex, but their
definition is relatively straightforward, once adic spaces have
been~defined.

\section{Lean and its Mathematical Library}
\label{sec:lean_mathlib}

Our formalisation of perfectoid spaces uses the Lean proof assistant
\cite{lean}, developed principally by Leonardo de Moura since 2013 at Microsoft
Research. Lean implements a version of the calculus of inductive constructions
\cite{cic}, with quotient types, non-cumulative universes, and proof
irrelevance.

In 2017, almost all the mathematical content of Lean's core library was split
off as a community-curated library called \mathlib \cite{mathlib-paper},
which is the basis for all mathematical developments in Lean, including our
perfectoid spaces formalisation.
Initially developed mostly by Mario Carneiro and Johannes Hölzl,
this library then gained many contributors,
including many people coming from mathematics departments.
Ultimately, everything described in this paper until
\cref{sec:valuations} included should be merged into \mathlib,
leaving only very specialised material in the perfectoid spaces project.
Some of our work was directly done inside \mathlib,
some was merged later, some is still to be merged.

The \mathlib library takes inspiration from many great efforts in other proof
assistants,
notably Isabelle, Coq, and Metamath,
but also has its own distinct flavour.
The most relevant comparison is with Mathematical Components
\cite{mathcomp}, since Lean's logical foundations are extremely close to those
of Coq.
On the mathematical side, the main difference is that \mathlib does not insist
on constructivity.
This makes it very appealing to regular mathematicians.
On the technical side, a very visible difference is \mathlib's heavy use of type
classes instead of
canonical structures as in \cite{mathcomp_structures}.
Both these mechanisms allow to handle extra structures on types.
For instance, for every type \verb+G+, \verb+add_group G+ is a type
packaging a composition law \verb$+$, a term \verb$0$ of type \verb$G$ and
properties like \verb$∀ g, g+0 = g$ etc.
The type family \verb+add_group+ is declared to be a \emph{type class}.
One can then declare \emph{instances} of that class.
They can be concrete instances,
like a group structure on $\ZZ$,
or composite instances,
such as the instance creating a group structure on $G \times H$ from a group
structure on $G$ and a group structure on $H$.
After setting up such a class and some instances, one can
declare that certain arguments of later definitions and lemmas should be
automatically inferred by searching the instance database, this is the instance
resolution procedure.

During the instance resolution procedure, Lean can only identify terms and types
that are definitionally equal, it does not use theorems asserting equalities.
This leads to encoding tricks that are already present in Coq\footnote{see e.g.
\cite[Section 3.2]{mathcomp_structures}} and Isabelle, for instance,
but are very counter-intuitive for mathematicians.
For instance, a metric space structure on some $X$ is usually defined as a
package containing a distance function from $X \times X$ to $\RR_{\geq0}$
and a bunch of conditions on this function.
Then one can define the topology associated to a metric space structure, and get
access to theorems proved about general topological spaces.
In \mathlib, and some other libraries, a metric structure on $X$ is%
\footnote{Actually what we describe is a simplified story.} a package
containing not only the distance function and its properties, but also a topology
and a property asserting that this topology is compatible with the distance
function.
There is a function which takes as input the traditional package and
outputs this ``augmented'' package, so defining concrete instances is no harder.
The difference matters with composite instances.
Say we start with two metric spaces $X$ and $Y$, and we want a metric structure
on $X \times Y$.
The traditional way would define a distance on $X \times Y$,
check the axioms, and then, after this
construction, prove a compatibility lemma saying the topology built from the
product metric equals the product topology.
But that equality will not be definitional,
and, in particular, this could block further instance resolution.
In the augmented packaging approach,
building the product instance involves defining the product distance,
feeding in the product topology, and using the mathematical content of the
compatibility lemma to provide the compatibility part of the package.
So the total amount of mathematical work is the same, but the metric topology on
$X \times Y$ is now definitionally equal to the product of the metric
topologies of~$X$~and~$Y$.

\section{Topology, Filters and Extensions}
\label{sec:topology_and_filters}

\subsection{Topology without Subsets}
\label{sub:topology_without_subsets}

Topology in \mathlib began when Johannes Hölzl ported his own Isabelle
work, documented in~\cite{isabelle_topology}.
The main traditional source for this is Bourbaki~\cite{bourbaki_tg_1_4}.
But the formal version goes quite a bit further in the systematic and functorial
use of filters and order relations,
as we will illustrate below.

Filters were introduced by Henri Cartan in order to unify various notions of
limits in topology.
They became standard in formalizations of general topology since
\cite{isabelle_topology}, see \cite{coquelicot,rouhling_phd}.
See also \cite{freek_filters} for an earlier use of (a variation on)
filters in a specialized elementary context.
A filter on $X$ is a set $\F$ of subsets of $X$ which
contains $X$, is stable under pairwise intersection, and such that every subset
containing an element of $\F$ belongs to $\F$.
The original definition was restricted to what we now call \emph{proper} filters,
those not containing the empty subset; the new definition is however more
functorial and hence more convenient.

Every point $x$ in a topological space $X$ gives rise to the filter $\nhds{x}$
of neighbourhoods of~$x$.
Complements of finite subsets in $\NN$ also form a filter~$\F_\NN$.
In $\RR$ one can also consider $\nhds{+\infty}$ made of
all subsets containing some $[A, + \infty)$, and similarly~$\nhds{-\infty}$.

Given a filter $\F$ on $X$ and a map $f \colon X \to Y$,
the pushforward filter $f_*\F$ on $Y$ consists of
subsets of $Y$ whose preimage under $f$ belong to $\F$.
The pullback $f^*\G$ of a filter $\G$ on $Y$ consists of subsets of $X$
containing the preimage of an element of $\G$.
For instance,
if $X$ is equipped with a topology induced from $Y$ by $f$ then
$\nhds{x} = f^*\nhds{f(x)}$.
Those operations are functorial:
$(f \circ g)_* = f_* \circ g_*$ and $(f \circ g)^* = g^* \circ f^*$.

A filter $\F$ is \emph{finer} than another one $\F'$ (on the same $X$) if
$\F' \subseteq \F$. We denote this order relation by $\F \leq \F'$.
Both mapping operations $f_*$ and $f^*$ are order preserving,
and they form a Galois connection:
$f_*\F \leq \G \iff \F \leq f^*\G$.

A filter $\G$ on some topological space $Y$ \emph{converges} to a point
$y \in Y$ if $\G \leq \nhds{y}$.
A function $f$ from $X$ to $Y$, \emph{tends to} a point $y$ with respect
to a filter $\F$ on $X$ if $f_*\F$ converges to $y$.
As announced this allows to unify the definition of limit of a sequence $u
\colon \NN \to Y$, a limit of a function $f \colon X \to Y$ at some point $x$ in
a topological space $X$, a limit of $f \colon \RR \to Y$ at $\pm\infty$, using
filters $\F_\NN$, $\nhds{x}$ and $\nhds{\pm\infty}$ respectively.
But, as noted in~\cite{isabelle_topology}, one can go one step further, consider
any filter $\F'$ on $Y$, and say that $f$ tends to $\F'$ with respect to $\F$ if
$f_*\F \leq \F'$.

Unpacking the definitions, one can check for instance that $f$ tends to $y$ with
respect to $\nhds{x}$ means that the preimage $f^{-1}(V)$ of any neighbourhood $V$ of
$y$ is a neighbourhood of~$x$.
But such an unpacking is often not needed,
and avoiding it leads to very efficient proofs by computation (in the
mathematical sense, not $\lambda$-calculus),
especially when combined with a systematic use of lattice operations on
the type of topologies or filters on a type.
For instance, given topologies $T$ and $S$ on $X$ and $Y$,
the product topology on $X \times Y$ is defined as
$\inf(\pr_1^*T, \pr_2^*S)$,
using the general notion of induced topology and infimum of topologies.
Similarly, products of filters are defined by
$\F \times \G = \inf(\pr_1^*\F, \pr_2^*\G)$,
and satisfy $f_*(\F \times \G) \leq f_*\F \times f_*\G$,
and $\nhds{(x, y)} = \nhds{x} \times \nhds{y}$.
None of these formulas explicitly feature subsets,
they operate at a higher level of abstraction.
As an example of using them, let us look at our proof
that if $f \colon Z \to X$ and $g \colon Z \to Y$ are continuous at $z_0$
then so is $h := \big(z \mapsto (f(z), g(z))\big)$.
In the Lean code below, \lstinline+map h 𝓕+ means $h_*\F$,
and \lstinline+×ᶠ+ is product of filters.
\begin{lstlisting}
lemma continuous_at.prod_mk {f : Z → X} {g : Z → Y}
{z₀ : Z} (h : continuous_at f z₀) (h' : continuous_at g z₀):
continuous_at (λ z, (f z, g z)) z₀ :=
calc map (λ z, (f z, g z)) (𝓝 z₀)
     ≤ (map f 𝓝 z₀) ×ᶠ (map g 𝓝 z₀)  : map_prod_mk _ _ _
... ≤ (𝓝 f z₀) ×ᶠ (𝓝 g z₀)               : prod_mono h h'
... = 𝓝 (f z₀, g z₀)                           : nhds_prod_eq.symm
\end{lstlisting}

This style of proof was strongly advocated by Hölzl,
who initiated that part of \mathlib.
It does not seem to be used in Isabelle or Coq.
In addition to a couple of sporadic lemmas,
like the above one,
we had to develop the theory of filter bases,
which is not so useful for the abstract story,
but very convenient to define concrete instances,
like the adic topology defined by an ideal $I$ in a ring $R$.
This $I$-adic topology is characterised by the fact that neighbourhoods of zero
are subsets containing a power of $I$.

\subsection{Extension by Continuity}
\label{sub:extension_by_continuity}

Extension by continuity is an important operation of elementary topology, and it
plays a crucial role in the completion of topological rings that we need in
order to define the structure presheaf on adic spectra. The following is
\cite[Theorem I.8.5.1]{bourbaki_tg_1_4} (recall that a topological space is
regular if every point has a basis of closed neighbourhoods).

\begin{theorem}
  \label{bourbaki_extension}
  Let $X$ be a topological space, $A$ a dense subset of $X$, and $f \colon A \to Y$
  a mapping of $A$ into a regular space~$Y$. A~necessary and sufficient
  condition for $f$ to extend to a continuous mapping $\bar f \colon X \to Y$ is
  that, for each $x \in X$, $f(y)$ tends to a limit in $Y$ when $y$ tends to $x$
  while remaining in $A$. The continuous extension $\bar f$ of $f$ to $X$ is
  then unique.
\end{theorem}

From a type theory point of view, we immediately spot a little difficulty here
since we need every function to be total on its domain type.
One way to fix this is to make the inclusion map $\iota \colon A \to X$ more
visible.
It is actually already present since ``$y$ tends to $x$ while remaining in $A$''
means that we use the filter $\iota^*\nhds{x}$.
So the extension problem becomes a factoring problem, i.e. finding $\bar f$ such
that $\bar f \circ \iota = f$.

This little perspective shift allows to consider the case where $\iota$ is not injective.
As we will explain, this generalisation turns out to be useful
in the theory of uniform spaces (in particular topological groups and rings).
We simply assume that $A$ and $X$ are topological spaces,
$\iota$ pulls back the topology of $X$ onto the topology of $A$, and the image
$\iota(A)$ is dense in~$X$.

Another issue related to totality of functions is that
\cref{bourbaki_extension} is too global.
We want a version that applies to functions that are not everywhere continuous.
The key example is inversion on a topological field.
In Lean, inversion is a total function,
the inverse of zero being defined as zero.
Of course we don't expect inversion to be continuous at zero,
but we still want to use some version of \cref{bourbaki_extension} in order to
define inversion on $\RR$ out of inversion on $\QQ$ for instance.
The traditional workaround is to apply the theorem to $\QQ \setminus \{0\}$
equipped with the induced topology.
We could do that in Lean,
but it creates some overhead which is more difficult to ignore than on paper,
and tends to pile up.

One last formalisation detail is that we do not want to carry around the limit
assumption each time we mention the extension~$\bar f$. We want an extension
operator $E_\iota$ transforming functions from $A$ to $Y$ into functions from
$X$ to~$Y$.
Of course the result won't have any useful properties if $f$ is awful. We can do
that using the axiom of choice. The definition of $E_\iota f(x)$ is: if
there exists some $y$ such that $f_*\iota^*\nhds{x} \leq \nhds{y}$ then choose
such a $y$ as $f(x)$, otherwise choose any element of~$Y$.
Our final version of the extension theorem is then:

\begin{theorem}
  \label{lean_extension}
  Let $\iota \colon A \to X$ be a map between topological spaces whose image is
  dense, and which pulls back the topology of $X$ to the topology of~$A$.
  Let $f$ be a map from $A$ to a regular topological space~$Y$.
  \begin{itemize}
    \item
      If $f$ is continuous at $a$ then $E_\iota f(\iota(a)) = f(a)$.
    \item
      Let $x$ be a point in $X$. If, for every $x'$ in a neighbourhood of $x$,
      there exists $y$ in $Y$ such that $f_*\iota^*\nhds{x'} \leq \nhds{y}$ then
      $E_\iota f$ is continuous at $x$.
  \end{itemize}
\end{theorem}

Note that the proof is almost exactly the same as the proof of
\cref{bourbaki_extension}, which uses injectivity of $\iota$ only for
psychological purposes. And locality also comes essentially
for free in the proof since continuity is a local property.

\Cref{bourbaki_extension,lean_extension} have an obvious drawback:
the existence of a limit assumption looks fairly difficult to check.
In undergraduate accounts of this theorem, it is replaced by a uniform
continuity assumption on $f$ coupled with a completeness assumption on the
target space~$Y$.
In such accounts both assumptions seem to require using metric spaces, which is
too restrictive for our purposes.
The next section addresses this problem by introducing uniform spaces.
This will allow us to go beyond metric spaces,
but won't solve the issue that uniform continuity is too much to ask for.
For instance, even in the most elementary case of extending multiplication from
$\QQ \times \QQ \to \QQ$ to $\RR \times \RR \to \RR$ we hit the problem that
this map is not uniformly continuous. In this case it is locally uniformly
continuous, but this won't happen for general topological rings.
A very good introduction to the next two sections is
\cite{rob_padic}, where Lewis explains his design decisions for the definition
of~$\QQ_p$, with its valued field structure%
\footnote{There is also a formalisation of $\QQ_p$ in \cite{univalent_p_adic}
from a purely algebraic point of view, which is irrelevant here}.
He is able to take several shortcuts
that are not available to us
in the more general setting of Huber rings and adic spaces.

\section{Completions}
\label{sec:completions}

\subsection{Uniform Notions}
\label{sub:uniform_notions}

Uniform structures are refinements of topological structures introduced by André
Weil as a common generalisation of metric spaces and topological groups, which
allow to discuss uniform continuity (hence the name) and completeness.
A \emph{uniform structure} on $X$ is a filter $\U$ on $X \times X$ satisfying some
conditions.
To each metric space $(X, d)$, we can associate the uniform structure~$\U$
generated by sets of the form $\{(x,y) \mid d(x,y) < \varepsilon\}$
for some positive~$\varepsilon$.
We include this example in order to relate uniform spaces to
metric spaces,
but it doesn't play any role in our formalisation.
A relevant example associates to each topological abelian group $G$, with
subtraction map $\delta \colon G \times G \to G$, the uniform structure
$\delta^*\nhds{0}$.
Here we assume $G$ is commutative for simplicity.
In our applications it will always be the underlying additive group of a topological ring.

Elements of a uniform structure~$\U$ are called entourages.
Two moving points $x$ and $y$ are declared arbitrarily close to each other if
the pair $(x, y)$ belongs to an arbitrary entourage.
The first condition imposed on $\U$ is that all entourages contain the diagonal.
This is analogous to the reflexivity condition in the definition of a distance
function.
The other conditions are analogous to the symmetry and triangle inequality
conditions.
However, no analogue of the separation axiom $d(x, y) = 0 \implies x = y$ is
enforced, allowing for non \separated spaces.
One can prove that, given a uniform structure $\U$, there is a unique topology
$T_\U$ on $X$ such that $\nhds{x} = (y \mapsto (x, y))^*\U$.
This topology is \separated if and only if the diagonal is the only subset of
$X \times X$ contained in all entourages.
In \mathlib, the topology is actually part of the uniform structure,
together with an axiom equivalent to the neighbourhood formula.
This is an encoding trick fully analogous to what we explained
in \cref{sec:lean_mathlib}.

Uniform structures can be pulled back: one can check that, for every map
$f \colon X \to Y$ and every uniform structure $\U$ on $Y$, the filter
$(f \times f)^*\U$ is a uniform structure on $X$ (however pushforward does not
work in general, the analogue of the triangle inequality is not preserved).
One can then define uniform continuity for maps $f$ between uniform spaces
$X$ and $Y$ by $\U_X \leq (f \times f)^*\U_Y$. Again this way of writing things
is slightly more abstract than what is usually seen in textbooks. For instance,
Bourbaki writes: $f \colon X \to Y$ is uniformly continuous if, for each
entourage $V$ of $Y$, there is an entourage $W$ of $X$ such that
the relation $(x, x') \in W$ implies $(f(x), f(x')) \in V$.
Bourbaki then writes that an immediate consequence of the definitions is that
uniformly continuous functions are continuous (for the underlying topologies).
But thinking in more abstract terms actually allows to break the proofs into
small lemmas that are needed anyway, and gives a computational proof which does
not mention entourages at all.
The lemmas assert that taking the topology underlying a uniform structure is
order preserving and ``commutes with pull-back'' (the later being called the
``commutativity property'' until the end of this paragraph).
The proof of continuity then becomes:
$T_{\U_X} \leq T_{(f \times f)^*\U_Y} = f^*T_{\U_Y}$
where the inequality comes from the order preserving property of $T$
applied to the uniform continuity assumption,
and the equality is the commutativity property.
The resulting inequality $T_{\U_X} \leq f^*T_{\U_Y}$ is
equivalent to continuity of~$f$.
In \mathlib, the commutativity property is definitional,
by the encoding trick, so the proof does not even need to mention it,
and ends up being only twice as many characters long as the sentence
``This follows immediately from definitions''.
\looseness=1

Similarly, one can extend the definition of Cauchy sequences to uniform spaces.
However this notion is not powerful enough when sequential continuity does not
imply continuity.
The solution here is to define a notion of Cauchy filters, not necessarily
coming from sequences.
A filter $\F$ of $X$ is Cauchy if it is proper and if $\F \times \F \leq \U$.
This indeed generalises the notion of Cauchy
sequences: a sequence $u \colon \NN \to X$ into a metric space or a topological group
is Cauchy if and only if the filter $u_*\F_\NN$ is Cauchy (in \mathlib, Cauchy
sequences are actually defined like this, but the elementary definition is then
proved to be equivalent).
By definition, a uniform space is complete if every Cauchy filter
converges to some point.
This is equivalent to the elementary
notion in the case of metric spaces.

Uniform spaces, uniform continuous functions and Cauchy filters were formalised
before in Isabelle/HOL and Coq \cite{coquelicot, rouhling_phd}.
The Coq formalisations currently use
a slightly awkward definition of uniform spaces,
but \cite{rouhling_phd} mentions plans to switch to the one explained above.
As far as we know,
everything from this point to the end of the paper has never been formalised
before.

\subsection{The Completion Functor}
\label{sub:the_completion_functor}

As explained in \cref{sec:background_and_overview},
completing topological rings (endowed with their uniform structure)
is the required abstraction of moving from polynomials
to general analytic functions.
It happens that \separated completions always exist for uniform spaces.
More precisely, the inclusion functor from the category of complete \separated
uniform spaces into the category of all uniform spaces admits a left adjoint.
The key part of that assertion is the following theorem:
\begin{theorem}[{\cite[Theorem~II.3.7.3 and Definition~II.3.7.4]{bourbaki_tg_1_4}}]
  \label{exists_completion}
  For each uniform space $X$, there is a complete \separated uniform space $\hat
  X$ and a uniformly continuous map $\iota_X \colon X \to \hat X$ with the following
  universal property: every uniformly continuous map~$f$ from $X$ to a complete
  \separated uniform space $Y$ uniquely factors through $\iota_X$.
  The space $\hat X$ is called the \separated completion of $X$.
\end{theorem}
One can apply this universal property to $\iota_Y \circ f$ when $Y$ is not complete and
\separated to lift $f$ to a morphism $\hat f \colon \hat X \to \hat Y$.
Uniqueness in the universal property then proves this lifting operation respects
composition, hence we indeed have a completion functor.

Bourbaki constructs $\hat X$ as the space of maximal%
\footnote{Bourbaki writes ``minimal'' because his order on filters is opposite to
ours.}
Cauchy filters on $X$,
and the map $\iota_X$ sends $x$ to $\nhds{x}$.
The first main task is to endow this space with a uniform structure which
$\iota_X$ pulls back to $\U_X$, and prove that $\iota_X$ has dense range.
The next task is to build factorisations.
One wants to use extension by continuity,
using completeness of the target space to provide the necessary
limits.
But $\iota_X$ cannot be injective if $X$ is not \separated.
Indeed any uniformly continuous map into a separated space sends non-separated
pairs of point into the same point.
So Bourbaki needs to first factor through the separated space $\iota_X(X)$,
and then apply \cref{bourbaki_extension}.
With our \cref{lean_extension}, this detour is not necessary.

In \mathlib, we used a variation on this construction,
suggested in \cite{james} and whose formalisation was started by Hölzl.
One could wonder how to relate this construction to the Bourbaki construction.
More seriously, there are more elementary constructions of completions
in the special case of metric spaces.
The \mathlib library constructs real numbers and $p$-adic numbers using
equivalence classes of Cauchy sequences of rational numbers.
We certainly want to make sure these are isomorphic to our general completions,
if only as a sanity check (note Bourbaki solves this problem
by avoiding any mention of the elementary approach).
We introduce the following definition. Given a uniform space $X$, we say that
$(Y, \iota \colon X \to Y)$ is an \emph{abstract completion of $X$} if $Y$ is a
complete \separated space, $\iota$ has dense image and pulls back the uniform
structure of $Y$ onto the uniform structure of $X$.
We then prove that this property is sufficient to ensure the expected universal
property, without any reference to the construction of $Y$, which is left
completely unspecified.
The usual abstract arguments then gives an isomorphism between any
two abstract completions of a given uniform space.
We can then check that both our general construction and the Cauchy sequences
completion of rational numbers are abstract completions,
hence isomorphic.

\subsection{Completing Groups and Rings}
\label{sub:completing_groups_and_ring}

As explained above, topological groups have a natural uniform structure.
Mathematically it sounds natural to
define a type class instance of \verb+uniform_space G+ from an instance of
\verb+topological_add_group G+. One can then use the completion
construction from the preceding section. Extending the group operations
to the completion $\hat G$ is immediate using the completion functor, since
these operations are uniformly continuous. The extensions are uniformly
continuous hence continuous, and $\hat G$ is also a topological group.
But this leads to a subtle issue: we now have two different uniform structures
on $\hat G$. The first one comes from the completion functor applied to the
uniform structure coming from the topological group structure on $G$.
The second one comes from the topological group structure built on $\hat G$.
Those two uniform structures are equal,
but this is a theorem which is not obvious in any way.
Bourbaki does prove this result, but the proof is hidden in the middle of a
discussion culminating in the following statement (simplified using our standing
commutativity assumption), to be compared with their earlier statement of
\cref{exists_completion}:
\begin{theorem}
  \label{bourbaki_group_completion}
  Any \separated topological abelian group $G$ is isomorphic to a dense
  subgroup of a complete group $\hat G$. The complete group $\hat G$ (which is
  called the completion of $G$) is unique (up to isomorphism).
\end{theorem}
In particular this theorem does not discuss at all whether the hat notation
denotes the same operation as in \cref{exists_completion}.
In this case it denotes the same \emph{set}, and the ambiguity about the uniform
structure is not important since they prove both constructions give the same
structure.
The situation deteriorates once we notice the separation assumption (which does
not appear in \cref{exists_completion}).
After proving \cref{bourbaki_group_completion}, Bourbaki
writes\footnote{This is slightly edited for simplification purposes.}:
Let $G$ be a group which is not necessarily \separated, and let
$G' := G/\overline{\{0\}}$ be the \separated group associated with $G$.
The completion of $G'$ is called the \separated completion of $G$, and denoted
by $\hat G$.
So we now have a third meaning of the hat notation, still very closely
related, but definitely \emph{not} the same set this time.
Bourbaki then states the universal property of $\hat G$, factoring morphisms in
the category of topological groups.

All this is pretty reasonable from a mathematical point of view, because all
three variants of the completion $\hat G$ are isomorphic.
And we want the formalisation to look just as friendly as the real thing, so
some thinking is needed.
On top of the fact that a formal definition needs to be one definition rather
than three definitions, there is a difficulty coming from the type class
mechanism.
Whatever the definition of \verb+completion G+, Lean needs to find a uniform
structure instance, and a topological group instance, without entering an
infinite loop, and without choosing an instance that later lemmas won't
automatically recognise as the right one.
Typically a lemma needing completeness of \verb+completion G+ could fail to
derive this automatically if Lean infers a uniform structure from the
topological group structure,
whether it will definitely succeed if it gets it from the completion
construction (whose completeness is part of the conclusion of
\cref{exists_completion}).

We already explained part of the solution: using \cref{lean_extension}
instead of \cref{bourbaki_extension}, we can completely forget about the
separation assumption in \cref{bourbaki_group_completion}.
The only trade-off is that we must give up the word subgroup since $G \to \hat
G$ will be injective only if $G$ is \separated.
The other part of the solution is to give up the idea that we have two
operations (building the uniform structure on a topological group and completing
a uniform space) whose commutation should be proved and then silently used.
Instead we introduce a predicate asserting compatibility of a uniform structure
and a group structure.
We then have a first lemma asserting that, from a topological group one can
build a uniform structure whose underlying topology is definitionally equal to
the original topology, and definitionally satisfies the compatibility predicate.
And another lemma saying that if a group structure on $G$ is compatible with a
uniform structure $\U$ then, on the completion of $G$, defined using only $\U$,
the group structure obtained by extension is compatible with the completed
uniform structure.
This last lemma has exactly the same non-trivial mathematical content
as in the commutation assertion.
But this repackaging avoids all the notation overloading and is
formalisation-friendly.

Once completions of abelian groups work nicely, the next task is to extend
multiplication on the completion of a topological ring.
As mentioned earlier, multiplication is not uniformly continuous, even in the
elementary cases.
So one must really check the limit assumption in \cref{lean_extension}, and the
proof is pretty subtle.
Bourbaki builds this extension in two steps because of the injectivity
assumption hidden in \cref{bourbaki_extension}.
In particular the meaning of the hat notation is not quite the same as in their
general discussion of completions.
Our formalisation does it in one step and proves, without silently
changing the meaning of notations:

\begin{theorem}
  Let $R$ be a topological ring, equipped with its natural uniform structure.
  There is a ring structure on the \separated completion $\hat R$ which is
  compatible with its topological structure, and such that $\iota_R$ is a ring
  morphism. If $f \colon R \to S$ is a continuous ring morphism between topological
  rings then $\hat f \colon \hat R \to \hat S$ is also a ring morphism.
\end{theorem}

Before realising that \cref{lean_extension} holds, we formalised Bourbaki's two
step completion of topological rings, and encountered some formalisation
difficulty.
Briefly, the general theory of uniform spaces give us a minimal equivalence
relation with a separated quotient.
For topological groups, one can prove $x \sim y$ if and only if $x - y$ is in
the closure of the singleton zero.
But this has mathematical content, and is certainly not definitional.
So the topology library endows this quotient with a uniform structure,
the algebraic theory endows it with a group structure,
but these structures live on different quotient \emph{types}.
One can think of ways to solve that problem more elegantly
than we did,
but we didn't pursue our investigations in this direction since, as explained above,
we found a way to completely bypass this issue.

The main point of this part of the story is that ordinary mathematics sets up
topological ring completions by silently identifying different isomorphic
mathematical objects.
Our annoyance while trying to formalise this led us to streamline the theory by
proving \cref{lean_extension}
and introducing abstract completions.

\subsection{Field Completions}
\label{sub:field_completions}

We also need topological fields,
i.e. topological rings $K$ where each non-zero has an inverse,
and inversion is continuous from $K^* := K \setminus\{0\}$ to
itself.
Complications arise because their completions, endowed with their ring structure
built in the previous section, can have zero divisors. Bourbaki gives the
following criterion, whose proof is basically left to the reader (the only hint
being to use \cref{bourbaki_extension}).

\begin{theorem}[{\cite[Proposition~III.6.8.7]{bourbaki_tg_1_4}}]
  \label{field_completion}
  The completion of a \separated topological field is a topological field if
  (and only if) the image under the mapping $x \mapsto x^{-1}$ of every Cauchy
  filter which does not have a cluster point at zero, is a Cauchy filter.
\end{theorem}

The separation assumption is not a restriction because a topological field is
either \separated or has the indiscrete topology.
In the later case the \separated completion is the zero ring, which is obviously
not a field.
In the formalisation, the first question is how to handle the condition that
inversion is continuous from $K^*$ to itself.
This definition implicitly equips $K^*$ with the induced topology.
Initially we used $K^*$ as a subtype, mapping to $K$ by inclusion and equipped
with the induced topology, and induced uniform structure.
This was manageable and we were able to prove the completion criterion, but
every step was slightly painful, because inclusions are harder to ignore in
formalised type theory.

On the algebraic side, it was decided long ago that manipulating inversion
didn't deserve a subtype.
Inversion is defined on $K$, and the definition of a field requires $0^{-1} = 0$.
For the purpose of this discussion, let us denote by $\inv$ this
weird function from $K$ to itself.
We can then require that, in a topological field, $\inv$ is continuous at every
non-zero point, and this is equivalent to the previous definition.
In the proof of \cref{field_completion}, we apply the extension result,
\cref{lean_extension}, to $\iota_K \circ \inv$.
Now the behaviour of $E_\iota(\iota_K \circ \inv)$ at zero is pretty mysterious,
since it uses the axiom of choice to pick an element of $\hat K$.
So we define $\hatinv \colon \hat K \to \hat K$ as mapping $0$ to $0$, and use
$E_\iota(\iota_K \circ \inv)$ elsewhere.
Modulo this little trifle at $0$, everything proceeds very smoothly.
Because \cref{lean_extension} is local, and $K^*$ is a neighbourhood of each of
its elements (from the separation assumption), we don't need to care about zero
when proving continuity of $\hatinv$ at non-zero points of $\hat K$.
We defined a type class \verb+completable_top_field+ which records the
assumptions of \cref{field_completion}, in order for the type class system to
automatically derive the field structures we will need.
It remains to provide the relevant instances of
\verb+completable_top_field+, but this is a story for the next section.

\section{Valuations}
\label{sec:valuations}
\subsection{Definition and Target Monoids}
\label{sub:definition_and_target_monoids}

A totally ordered group (resp. totally ordered monoid) is a commutative group
(resp. commutative monoid) $\Gamma$, whose composition law is written multiplicatively,
together with a total order such that the multiplication map $\gamma' \mapsto
\gamma\gamma'$ is order preserving for every $\gamma$ in $\Gamma$. To every
totally ordered group $\Gamma$ one can associate a totally ordered monoid
$\Gamma_0 := \Gamma \cup \{0\}$ where the multiplication and order relation are
extended by specifying that $0\gamma = 0 \times 0 = 0$ and $0 \leq \gamma$ for
every $\gamma$ in $\Gamma$.

A valuation on a ring $R$ is a map $v$ from $R$ to $\Gamma_0$ for some totally
ordered group $\Gamma$ such that:
\begin{itemize}
  \item $v(0) = 0$ and $v(1) = 1$
  \item $v(xy) = v(x)v(y)$ for all $x$ and $y$
  \item $v(x+y) \leq \max\big(v(x), v(y)\big)$ for all $x$ and $y$.
\end{itemize}
This terminology is a bit unfortunate, and not consistent with
\cite{bourbaki_ac_1_7}, but it is systematically used in the adic spaces
literature, although a valuation in this sense is closer to a norm or a seminorm than a
traditional valuation.

Notice how, once again, the above definition focuses on the construction of
$\Gamma_0$, starting from an ordered group and adding an element, rather than
the properties of the resulting object.
The first consequence of this psychological trick is to help to hide the use of
many formal properties of $\Gamma_0$.
For instance we need at some point that, for every $x$, $y$ and $z$ in
$\Gamma_0$, $xz < yz$ implies $x < y$.
Such kind of results are not even stated in \cite{bourbaki_ac_1_7}, and the
systematic proof is extremely boring.
There is a case split according to whether $x$, $y$ and $z$ belong to the original
$\Gamma$ or are $0$,
so there are eight possibilities,
but only the case where all elements are non-zero is actually relevant, and
needs a lemma about totally ordered groups.

Initially we formalised this story using the type constructor
\verb+with_zero Γ+ which is simply a wrapper for \verb+option Γ+ but has all the
relevant type class instances (assuming \verb+Γ+ has its own relevant instances).
We used Lean's coercion mechanism to automatically insert the ``inclusion'' map
from $\Gamma$ to $\Gamma_0$. In set theory, this is indeed a true
inclusion, but there are no inclusions between types in type theory
as a term can only have one type.
The \verb+norm_cast+ tactic,
by Paul-Nicolas Madeleine \cite{norm_cast_report},
greatly alleviates the pain of invoking lemmas about such coercions.

We then addressed the proof issue by taking advantage of how convenient it is to
write specialised easy automation in Lean.
Our tactic \verb+with_zero_cases+ made the case distinction and
dispelled all irrelevant cases using Lean's simplifier and \verb+norm_cast+,
without any human assistance, leaving only the relevant case to prove by hand.

However that does not solve the following discrepancy.
In an elementary context, valuations take value in $\RR_{\geq 0}$
or $\QQ_{\geq 0}$.
Set-theoretically, it is almost true that
$\RR_{\geq 0} = (\RR_{> 0})_0$.
This tiny lie only requires that we forget that the $0$ in $(\RR_{> 0})_0$ is meant
to be an extra element coming from nowhere in particular, hence has nothing to
do with the neutral element of the additive group $\RR$.
It gets harder to ignore with type theoretic foundations.
It means that merging the elementary theory with our abstract theory would require
non-trivial glue.
So we stepped back and gave up using \verb+with_zero+.
We defined the concept of totally ordered commutative monoid with zero as a type
equipped with a composition law, a total order, and special elements $0$ and $1$
with enough properties to guarantee that all instances are isomorphic (as
ordered commutative monoids) to some $\Gamma_0$.
This corresponds to a type class which admits both
\verb+with_zero Γ+ and \verb+{x : ℝ // 0 ≤ x}+ as instances.
Once again, focusing on properties instead of constructions gives us the needed
extra flexibility at no cost.

\subsection{Valuation Topology and Valuation Extensions}
\label{sec:valuation_topology_and_extensions}

In order to construct a valuation on stalks  in
\cref{sec:adic_spectrum_and_adic_spaces},
we will start with a field equipped with a valuation $v$,
endow that field with a topology associated to $v$,
complete, and extend $v$ to the completion.

\hyphenation{to-po-lo-gy}
A valuation $v$ on a ring $R$ with values in $\Gamma_0$ defines a topology
characterised by the fact that, for every $x$ in $R$, neighbourhoods of $x$ are
subsets containing $\{y \mid v (y - x) < \gamma\}$ for
some $\gamma \in \Gamma$
(existence of this topology is not obvious, it needs to be proven).
The first main result of this section is:

\begin{proposition}
  The topology coming from a valuation on a ring makes it a non-Archimedean
  topological ring.
  If this ring is (commutative and) a field, then it is completable,
  and the valuation extends to the completion, with the same target monoid.
\end{proposition}

In this statement, non-Archimedean means having a fundamental system of
neighbourhoods of zero consisting of additive subgroups.
We now focus on the extension part of the proposition, for a field $K$.
The elementary example is $K = \QQ$ equipped with some $p$-adic valuation, and
the extension is the $p$-adic valuation on $\QQ_p$.

The proof in \cite[VI.5.3]{bourbaki_ac_1_7} is extremely short. Translated into our
notation, it says: According to Proposition 1, $v|_{K^*}$ is a continuous
representation of $K^*$ into $\Gamma$ (equipped with the discrete topology),
hence can be extended to a continuous representation $\hat v$ of $\hat K^*$ into
$\Gamma$. The relation $\hat v(x + y) \le \max\big(\hat v(x), \hat v(y)\big)$
holds in $K^*$, hence stays true in $\hat K^*$ by continuity.
Hence $\hat v$, extended by $\hat v(0) = 0$, is a valuation.

At first reading, we thought the first ``hence'' was implicitly referring to the
extension promised by \cref{exists_completion}. But this would require
identifying $\hat K^*$ with the uniform completion of the multiplicative
group $K^*$. However the uniform structure induced by $K$ on $K^*$ is not the
uniform structure associated to the topological group $(K^*, \times)$, otherwise
multiplication would be uniformly continuous. After exchanging a couple of
emails with current and former collaborators of Nicolas Bourbaki, it looks like
Bourbaki meant to apply \cite[III.3.3.5]{bourbaki_tg_1_4}. In the meantime, we
formalised another proof.

Before sketching this alternative proof, let us consider the
non-Archimedean triangle inequality argument in the above proof.
It looks rather uncontroversial until one notices that $x + y$ can land outside
$K^*$ (or $\hat K^*$) even if $x$ and $y$ are both in $K^*$ (or $\hat K^*$).
This is a bit embarrassing, but not a serious difficulty:
one can separately argue that if $x + y = 0$ in $\hat K$ then the inequality
holds for free.
But this will complicate the formal proof, which also requires spelling out
other details.

Fortunately, all this is avoided by our alternative proof. We endow $\Gamma_0$
with a topology where non-zero points are open but neighbourhoods of zero are
subsets containing some open interval $\{\gamma \mid \gamma < \gamma_0\}$
(this is not an exotic topology to put on $\Gamma_0$,
but it seems Bourbaki tried to avoid introducing it).
We then apply \cref{lean_extension} to the full map $v \colon K \to \Gamma_0$.
Proving the assumptions of the theorem involves, as in the
elementary proofs, arguing that the valuation is locally constant on $K$. Of
course we also need to prove that $\Gamma_0$ is a regular topological space and
its order relation plays nicely with the topology.

The conclusion is always the same: in order to avoid formalisation pain when
filling in details or fixing little mistakes,
we used a slightly different point of view which turns out to be more natural.
Indeed one cannot extend a valuation by seeing it purely as a multiplicative
group morphism and hope that it will play nicely with the triangle inequality,
which is about addition.

\section{Presheaves and Sheaves}
\label{sec:sheaves}

A presheaf $\F$ on a topological space $X$ with values in a category $\C$
associates an object $\F(U)$ of $\C$ to each open subset $U$ of $X$,
and a so-called restriction morphism $\res_{U, V} \colon \F(U) \to \F(V)$
whenever $V \subset U$,
with $\res_{U, W} = \res_{U, V} \circ \res_{V, W}$ when $W \subset V \subset U$.
When $\C$ is a concrete category, as in all our examples,
elements of $\F(U)$ are called sections of $\F$ on~$U$.
The most elementary examples are sheaves of functions,
for instance $\F(U)$ could be the ring of real-valued continuous functions on
$U$, with the usual restriction operation.
In our context, the structure sheaves of adic spaces will be thought as
sheaves of analytic functions,
but their definition will involve nothing like functions
(see \cref{thm:spa} below).
In particular building restriction morphisms, and checking their compatibility,
will involve abstract work.

In order to reasonably think of sections of a presheaf as functions,
they must satisfy some locality and gluing conditions.
One then says that $\F$ is a sheaf.
Roughly, for each open covering $(U_i)$ of an open set $U$,
and each collection of sections $s_i \in \F(U_i)$ that are compatible,
i.e. $\res_{U_i, U_i \cap U_j} s_i = \res_{U_j, U_i \cap U_j} s_j$
for all $i$ and $j$,
there is a unique $s \in\F(U)$ which restricts to $s_i$ on each $U_i$.
The precise definition more carefully uses the full structure of $\C$.
In our case, $\C$ is the category of complete \separated topological rings,
and the bijection between $\F(U)$ and the collections of compatible sections
$(s_i)$ must be a homeomorphism.

As is often the case, we will construct sheaves by first defining them
on elements of a basis $\mathcal{B}$ of open subsets,
and then we extend by setting $\F(U) = \varprojlim_V \F(V)$,
the projective limit being over elements of $\mathcal{B}$ contained in $U$.
In practice the partially ordered set over which we took the projective limit
was slightly finer than the obvious one, however one can prove that the limit
we took is naturally isomorphic to the ``correct'' limit.
Because projective limits are only defined up to isomorphism,
this is not a problem.

We will also need local information encoded in stalks of $\F$ at a point $x$,
$\F_x = \varinjlim_{U \ni x} \F(U)$,
where the injective limit is over open subsets containing $x$.
When the sheaf is constructed from a basis,
one can prove it is sufficient to consider basic subsets containing $x$.
In our situation, we will first forget about the topologies on $\F(U)$ and
consider stalks of sheaves of rings.

Sheaves of rings were introduced in Lean during the formalisation of schemes
\cite{lean-schemes}.
We reused that formalisation, extending it to sheaves of \emph{topological}
rings.
This is an ad hoc formalisation which does not rely on \mathlib's
category theory library.
We found this to be easier for our purposes,
but it is not sustainable since turning to sheaves with values in other
categories would need a lot of code duplication.

\section{Adic Spectra}
\label{sec:adic_spectra}

Modulo the modifications explained above, all the mathematics in \cref{sec:topology_and_filters,sec:completions,sec:valuations}
was already present in \cite{bourbaki_tg_1_4, bourbaki_ac_1_7} going back to the
40s and 60s respectively.
Sheaves of topological rings are also from the same era.
We now jump to the 90s with the construction of $\Spa(A, A^+)$, the adic
spectrum of a Huber pair.
\looseness=-1

\subsection{Valuation Spectra and Continuous Valuations}
\label{sub:continuous_valuations}

The first step towards constructing $\Spa$ is
to consider the valuation spectrum $\Spv(R)$ of a ring $R$.
The informal definition starts by saying that two valuations
$v \colon R \to \Gamma_0$ and $w \colon R \to \Delta_0$ are \emph{equivalent} if
there exists an isomorphism $\varphi$ between the monoids $v(R)$
and $w(R)$ such that $w = \varphi \circ v$.
Then $\Spv(R)$ is defined as the ``set'' of ``equivalence classes'' of
valuations on $R$.
But there is no set of valuations, because the target monoids can be arbitrarily
large.
In type theory, this translates into universe issues.
Fortunately there are well-known reformulations that avoid these issues.
In particular, two valuations are equivalent if and only if they pull back the
order relation on their target monoids to the same preorder on $R$.
Since preorders on $R$ do form a set, one can define $\Spv(R)$ as the
subset of those coming from a valuation.
Since we still want to pretend $\Spv$ is defined as a quotient,
we then prove a series of lemmas which provides the usual quotient interface.

The next construction starts with a topological ring $R$.
Here there is a notion of a continuous valuation $v \colon R \to \Gamma_0$,
which is more subtle than asking for continuity when the target is equipped with
the topology discussed in \cref{sec:valuation_topology_and_extensions}.
It goes through replacing the valuation by some canonical representative of its
``equivalence class'' before checking continuity (one for which $\Gamma$ is as small as possible).
This enables us to define the subtype $\Cont(R) \subset \Spv(R)$ of
``continuous valuations'' on~$R$,
where the quotation marks highlight both the fact that equivalence classes of
valuations are now called valuations, and that they are not even equivalence
classes.

\subsection{\texorpdfstring{$\Spa(A, A^+)$}{Spa(A, A+)} and its Structure Presheaf}
\label{sub:Spa}

In discrete commutative algebra a large role is played by
Noetherian rings, because of their finiteness properties.
In the theory of adic spaces, there are no Noetherian assumptions.
However, the rings that show up are all endowed with a topology,
and one can recover finiteness constraints by
asking for a subtle interplay between the topological and algebraic structures.
\emph{Huber rings} form such a class of topological rings.

A topological ring $A$ is Huber if there exists some open subring $A_0$
whose topology is $I$-adic for some finitely generated ideal $I$
(the $I$-adic topology was defined at the end of
\cref{sub:topology_without_subsets}).
A Huber pair $(A, A^+)$ is a Huber ring~$A$ equipped with an open subring~$A^+$
which is integrally closed and power bounded (the definition of those conditions
will not be relevant below).
As a sanity check for our formalisation, we proved that $(\QQ_p, \ZZ_p)$ is
a Huber pair.
Doing that already reveals that all those definitions involving open subrings
are stated too rigidly for type theory.
Indeed \mathlib defines $\ZZ_p$ as a subtype, in order to equip it with all its
extra structures through the type class mechanism.
So we replaced open subrings by open embeddings of rings.

As a set,
the adic spectrum $\Spa(A,A^+)$ is defined to be
$\{ v \in \Cont(A) \mid \forall x \in A^+, v(x) \leq 1\}$.
We now set $X = \Spa(A, A^+)$ for some fixed Huber pair.
In order to put a topology on $X$,
to each $s$ in $A$ and each finite subset $T \subset A$ such that $T \cdot A$ is
open,
one associates the so-called rational open subset
$R(T/s) := \{v \in X \;|\; v(s)\neq 0 \land \forall t \in T, v(t) \leq v(s) \}$.
Here $T \cdot A$ is the additive subgroup generated by products of elements of
$T$ and general elements of $A$.
We also associate to each valuation $v$ on $A$ its support
$\supp(v) := \{x \in A \;|\; v(x) = 0 \}$.
This is a prime ideal, hence $A/\supp(v)$ has a field of fractions
$K_v$.
The valuation $v$ extends to $K_v$ where it defines
a topological field structure as in
\cref{sec:valuation_topology_and_extensions}.

\begin{theorem}
  \label{thm:spa}
  Let $(A, A^+)$ be a Huber pair, and set $X = \Spa(A, A^+)$.
  \begin{enumerate}
    \item
      The rational open subsets $R(T/s)$ form a topology basis on $X$.
    \item
      For each $(T, s)$ as above,
      there is a topology on the localisation $A[s^{-1}]$ turning it into a
      non-Archimedean topological ring,
      denoted by $A(T/s)$,
      which is universal for the following properties:
      the localisation map $\varphi \colon A \to A(T/s)$ is continuous,
      $\varphi(s)$ is invertible,
      and the image $(t \mapsto \varphi(t)\varphi(s)^{-1})(T)$ is
      power-bounded.
    \item
      There is a presheaf $\O$ of complete \separated topological
      rings on $X$ whose value on each $R(T/s)$ is naturally isomorphic to
      the \emph{\separated completion} $A\langle T/S\rangle$ of $A(T/s)$.
    \item
      For every $x$ in $X$, there is an equivalence class $v_x$
      of valuations on the stalk $\O[X, x]$ such that,
      for every $v$ representing $x$,
      the representatives of $v_x$ factor through $\widehat{K_v}$.
  \end{enumerate}
\end{theorem}

Explaining the proof of this theorem is far beyond the scope of this paper,
but we want to focus on some issues that arose during its formalisation,
and also indicate how the previous sections come together in the proof.
This will be done in
\cref{sub:Huber_localization,sub:presheaf,sub:the_stalk_valuation} below.

Before doing that, we need a disclaimer.
An important theorem in the theory,
necessary for any significant further work,
states that two natural topologies on $X$ coincide;
the topology on $X$ is thus defined by mathematicians to be ``both of them''.
The proof that the topologies coincide is
Theorem~3.5(ii) of~\cite{huber_valuations} or
Theorem~7.35(2) of~\cite{wedhorn}.
We did \emph{not} formalise the proof of this theorem.
The formal proof would probably involve months of work,
developing a good interface for
primary and secondary specialisations of valuations on rings,
as is done in Section~2 of~\cite{huber_valuations} and its references,
or sections~4 and~7 of~\cite{wedhorn}.
This means that we have to choose a topology for our definition, and
we choose the topology generated by the rational open subsets,
because this one is more convenient for us later.
The proof that our definition of the topology of an adic space
(\cref{sec:adic_spectrum_and_adic_spaces})
agrees with ``the definition in the literature''
thus may invoke a theorem in the literature
which we have not yet formalised, depending on what ``the'' definition
of the topology in the literature is.

We remark here that omitting the formalisation of this proof
causes us trouble again, when defining the presheaf on $\Spa(A,A^+)$
in \cref{thm:spa}(3) above. The definition we formalise
is easily checked using a non-formalised theorem to be the same as the
definition in the literature. See \cref{sub:presheaf} below for more details.

\subsection{Localisation, elaboration, and the community}
\label{sub:Huber_localization}

The first issue we want to discuss is that of balancing readable statements
against the demands of Lean's elaborator.
A typical lemma that shows up in the proof of the second item
of \cref{thm:spa} is the following.
\begin{lemma}
  Let $A$ be a Huber ring and let $T$ be a subset of $A$
  such that the ideal generated by $T$ is open.
  For every open subgroup $U$ of $A$, $T \cdot U$ is open.
\end{lemma}
\noindent
The next code block is our formalisation of this statement.
The last line is the conclusion and the different kind of parentheses and brackets
surrounding arguments indicate how they
will be provided when using the Lemma.
Regular parentheses flag arguments that will be given explicitly, curly
brackets are implicit arguments that will be deduced from the type of explicit
arguments, and square brackets will be deduced by instance resolution as explained in
Section~\ref{sec:lean_mathlib}.
\begin{lstlisting}
lemma mul_T_open {A : Type*} [Huber_ring A]
  {T : set A} (hT : is_open ((ideal.span T) : set A))
  (U : open_add_subgroup A) :
  is_open (↑(T • span ℤ (U : set A)) : set A) :=
\end{lstlisting}
This statement features several type ascriptions, an explicit coercion
operator \lstinline{↑}, and an ad-hoc coercion trick.
The coercion mechanism in Lean 3 works great in simple situations,
but not always when several instances must be resolved in the correct order.
In \mathlib, the type \lstinline{ideal A} bundles a subset of $A$
with the expected properties.
It has a coercion to \lstinline{set A}.
But in assumption \lstinline{hT}, Lean also needs to figure out
the topological space that \lstinline{is_open} refers to.
The type ascription helps it doing those things in the correct order.
Things pile up even more in the conclusion of the lemma,
where the (bundled) open subgroup $U$ is seen as a $\ZZ$-submodule by first
forgetting everything but the underlying subset and then taking the
$\ZZ$-span to build a (bundled) submodule whose underlying set is
exactly the same.
Lean then needs to figure out the meaning of \lstinline{•} which is the
overloaded notation for scalar multiplication (more on that instance below),
and then again find the relevant topological space.
Coercions to subsets are especially tricky because
\lstinline{set X} is defined as the function type \lstinline{X → Prop},
and coercions to functions receive a special treatment,
which can be misleading in the \lstinline{set} case.
Using the explicit coercion operator helps because coercions to
functions have a different dedicated operator.

In conclusion, we managed to write a somewhat recognisable statement,
but the situation is not perfect.
It may be improvable by more expert Lean users,
but we still claim this is not as easy as we'd like it to be.

The second issue is related to the first,
and has to do with the notation $T \cdot U$.
One of the benefits of introducing this notation
(and the corresponding instances of algebraic typeclasses)
is that it allows to use all sorts of results from the algebraic library:
submodules of an algebra form a semiring,
that is a semialgebra of the semiring of subsets
(under elementwise multiplication, followed by the linear span).

However, not all of these facts were available in \mathlib
at the time of our development.
The following instances were available (here $A$ is an $R$-algebra):
\lstinline{semigroup (submodule R A)},
\lstinline{comm_semiring (ideal R) }, and
\lstinline{algebra R R}.

In our project, we needed \lstinline{semiring (submodule R A)}.
But when we added this instance to the system,
type class search could then find two different instances
of \lstinline{semiring (ideal R)}, and they were not definitionally equal.
One came from the instance of \lstinline{comm_semiring (ideal R)} in \mathlib.
The other came from combining the two instances
\lstinline{semiring (submodule R R)} and \lstinline{algebra R R},
since \lstinline{ideal R} is defined to be \lstinline{submodule R R}.

The solution was to change \mathlib,
generalising the \lstinline{semigroup (submodule R A)} instance to a \lstinline{semiring}
and using this new instance to redefine
the existing instance of \lstinline{comm_semiring (ideal R)}.
This change was proposed in pull request \#856 to \mathlib.
As long as this pull request was not merged upstream,
making progress on localisations of Huber rings was not really possible.
This very unfamiliar experience for mathematicians was frustrating,
although it lasted only one week.

At the time the maintainers of \mathlib had very little time
for review and merging of pull requests.
Altogether this episode fostered reflection about
the processes involved in maintaining \mathlib.
Our overall experience is that maintenance is now smoother than before.
Nevertheless, this experience raises questions for the future.
Are lock-ups like the one explained above expected to occur more often
when libraries and formalisation efforts scale up?

\subsection{The Structure Presheaf}
\label{sub:presheaf}

We now comment on the third item of \cref{thm:spa}.
This was more problematic than it should have been,
because of our incomplete interface to the topological space structure on
$X$.
Recall that if $s\in A$ and $T\subset A$ is a finite subset such that $T\cdot A$
is open,
then there is a rational open subset $R(T/s)$ of $X$
defined by some inequalities.
We would like the ring of functions on $R(T/s)$
to be the completion $A\langle T/s\rangle$ of $A(T/s)$.
Because the rational open subsets form a basis for the topology
(by \cref{thm:spa}(1)),
a function on a general open set~$U$ should be determined by its
restrictions to the rational open subsets contained in~$U$.
So we would like to define $\O(U)$,
for $U$ a general open subset of $X$,
to be the subset of the product of the $A\langle T/s\rangle$
such that $R(T/s)\subseteq U$,
consisting of functions which ``agree on overlaps''.
There is a construction (Proposition~8.2 of~\cite{wedhorn})
which shows that if $R(T_1/s_1)\subseteq R(T_2/s_2)$ is an
inclusion of rational open subsets, then there is a natural
ring homomorphism $A\langle T_2/s_2\rangle \to A\langle T_1/s_1\rangle$,
which is an isomorphism if the inclusion is an equality.
We have not yet formalised the construction in this generality,
because it relies on our missing proof that two topologies
coincide. We avoided this issue by proving that
$R(T_1/s_1)\cap R(T_2/s_2)=R(T_1T_1/s_1s_2)$,
defining ring homomorphisms
$A\langle T_1/s_1\rangle\to A\langle T_1T_2/s_1s_2\rangle$
and
$A\langle T_2/s_2\rangle\to A\langle T_1T_2/s_1s_2\rangle$,
and defining $\O(U)$ to be
the elements $(f_i)$ of
the product of the $A\langle T_i/s_i\rangle$
for which $R(T_i/s_i)\subseteq U$,
such that the restrictions of $f_i$ and $f_j$
agreed in $A\langle T_iT_j\rangle/s_is_j$.
This definition is mathematically equivalent to the usual one,
although the obstruction to formalising the proof of this in Lean
is that we need primary and secondary specialisations of valuations,
which we do not yet have.
However it is not too difficult to show in Lean that our definition satisfies the
axioms of a presheaf, which is what we need.

\subsection{Stalk Valuations}
\label{sub:the_stalk_valuation}

Having defined the presheaf $\O$ on $X$,
the last piece of structure we need is a valuation on each stalk.
For a point~$x$ of $X$, let~$v$ be a valuation in its equivalence class.
The stalk of $\O$ at~$x$ is the colimit of the rings
$\O(U)$ as $U$ runs through the open subsets of $X$
containing $x$. Because the rational open subsets $R(T/s)$ form a basis
for the topology, this colimit is isomorphic to the colimit of the
rings $A\langle T/s\rangle$ where $R(T/s)$ runs through the rational
open subsets of~$X$ containing~$x$, so our task is to put a valuation
on this colimit which naturally extends~$v$. The universal property of the
colimit tells us that we can define a ring homomorphism from the colimit
by giving a collection of ring homomorphism from each $A\langle T/s\rangle$
which are compatible with our restriction maps. The valuation $v$ extends
to a valuation $\hat v$ on $K_v = \Frac(A/\supp(v))$ and, by
\cref{sec:valuation_topology_and_extensions}, to a valuation
on its completion. Thus it suffices to give
ring homomorphisms $A\langle T/s\rangle\to\widehat{K_v}$
for each pair $(T,s)$ such that $x\in R(T/s)$, which are compatible
with all restriction maps. The assumption $x\in R(T/s)$ implies that
$s \not\in \supp(v)$. Hence the map $A \to K_v$
extends to a ring homomorphism $A[s^{-1}]\to K_v$. The fact that
$x\in R(T/s)$ also implies that this homomorphism is continuous on $A(T/s)$
which is $A[s^{-1}]$ endowed with the topology coming from~$T$,
and we deduce the existence of a
ring homomorphism $A\langle T/s\rangle\to \widehat{K_v}$
using the completion functor.
It is
straightforward to check that these ring homomorphisms are compatible with our
restriction maps and hence they give rise to a ring homomorphism from
the stalk of $\O$ at~$x$ to $\widehat{K_v}$; composing
with the valuation $\hat v$ on $\widehat{K_v}$ gives us
the valuation we require.
\looseness=-1

In conclusion, to a Huber pair~$(A, A^+)$ we are able to associate
its adic spectrum $X = \Spa(A, A^+)$,
together with a presheaf of topological rings,
and a valuation on each stalk of the presheaf.

\section{Adic and Perfectoid Spaces}
\label{sec:adic_spectrum_and_adic_spaces}

An adic space is a topological space equipped with a sheaf of complete
topological rings with valuations on stalks which is locally isomorphic to
the adic spectrum of a Huber pair; see below for a formal definition.
These isomorphisms are taking place in an auxiliary
category which only plays a role in the definition of an adic
space; this category does not seem to be taken too seriously
by mathematicians and it does not even seem to have an established name.
References~\cite{huber_valuations} and~\cite{wedhorn} call it $\mathcal{V}$,
and we call it {\tt CLVRS}
(for ``complete locally valued ringed space'').
Its objects consist of the following data:
a topological space~$X$,
a sheaf of complete topological rings $\O[X]$ on~$X$,
and for each $x\in X$ an equivalence class of valuations $v_x$
on the stalk of $\O[X]$ at $x$,
whose support is the unique maximal ideal of $\O[X,x]$.
It is clear, although rarely explicitly mentioned,
that an open subspace of a topological space equipped with such structure,
naturally inherits such a structure.
We adopt the usual mathematical tradition of
conflating the object $(X,\O[X], (v_x)_{x\in X})$ of~$\mathcal{V}$
with its underlying topological space~$X$.
A morphism from the object $(X,\O[X],(v_x)_{x\in X})$
to $(Y,\O[Y], (v_y)_{y\in Y})$
consists of a continuous map $f \colon X\to Y$,
and a morphism of sheaves of topological rings
$\O[Y]\to f_*\O$
such that $v_x$ pulls back to $v_{f(x)}$ along the induced morphism
$\O[Y,f(x)]\to \O[X,x]$ of stalks. This gives
us a notion of isomorphism in this category.

In~\cite{wedhorn} we see a larger category
$\mathcal{V}^{\mathrm{pre}}$
where $\O$ is only assumed to be a presheaf
rather than a sheaf.
We used an even more general category
{\tt PreValuedRingedSpace},
and our a definition of an adic space is
an object of {\tt CLVRS} which is locally isomorphic
(in {\tt PreValuedRingedSpace})
to the continuous spectrum of a Huber pair:
\begin{lstlisting}[aboveskip=.2cm, belowskip=.2cm]
def is_adic_space (X : CLVRS) :=
∀ x, ∃ (U : opens X) (A : Huber_pair), x ∈ U ∧ (Spa A ≊ U)}
\end{lstlisting}
Since {\tt CLVRS} is a full subcategory of {\tt PreValuedRingedSpace},
this is equivalent to the usual definition.
So the last main piece of work is to set up those two categories.
Instantiating $\Spa(A, A^+)$ as an object of
{\tt PreValuedRingedSpace} is mostly restating things we did before.

Once the definition of an adic space is formalised,
it is not hard to formalise the definition of a perfectoid space;
it simply modifies the previous definition by asking that each
Huber pair is perfectoid.
Being perfectoid is just a relatively straightforward predicate
on a Huber pair, demanding essentially that there is a prime number~$p$
and an open subring $A_0$ of~$A$ such that $x\mapsto x^p$
is a surjection from $A_0/p$ to itself,
together with the ``deep ramification'' condition mentioned in the introduction.
This condition forces every element of~$A_0$ to have a $p$th root modulo~$p$;
this algebraic predicate implies that many functions on the space
have $p$th roots which are also functions on the space, which
has non-trivial geometric implications.
Scholze's insight that this is an important condition cannot
be explained here, although its importance in modern mathematics
cannot be overstated.

However, we have almost nothing beyond the definition.
We have one example -- the empty perfectoid space.
Even this example involved some work,
boiling down to the statement that the unique map
from an infinite product of trivial topological rings
to the trivial topological ring
was an isomorphism;
this is what the sheaf axiom boils down to in this case.
We are close to examples where
the underlying space has one element,
but non-trivial examples will be harder.

\section{Future Work and Concluding Thoughts}

We have developed a robust theory of completion of topological groups and rings
in the Lean theorem prover.
We have proved universal properties which characterise our constructions.
We have tested the usability of our theory
by developing a theory of continuous valuations and adic spectra
on top of it (and enough sheaf theory).
We developed the theory of adic spectra far enough to be able to
formalise the definitions of adic and perfectoid spaces.
At two points our construction differs slightly from the construction in the literature,
however standard (unformalised) theorems in the literature
can be used to show that
our nonstandard definitions are the same as the standard ones.
Very natural future work would be
to develop the theory of primary and secondary specialisations
of valuations,
which would allow to start developing properties of
adic and perfectoid spaces,
and would also enable us to formally prove these theorems.

Part of our experiment was to see if Lean could handle complex mathematical
objects without timing out, or running out of memory, or similar issues that do
not appear at all in informal mathematics.
Another part of the experiment was that we set out to formalise a definition
that brought together several different parts of Lean's mathematical library.
These parts were not designed with a formalisation of perfectoid spaces in mind,
but we could still successfully make all the different theories cooperate.
We feel that, since a proof assistant can handle the definition of a perfectoid space,
then they should be able to handle general modern mathematical definitions.

Such modern definitions are uncommon in the formalisation world,
perhaps because so few mathematicians know how to use formalisation software.
This is a sad state of affairs which we hope will change.
Hopefully our contribution will help to clear up the misconceptions that proof
assistants are only usable by people with a computer science degree, or can
only prove theorems about undergraduate-level objects (as in \cite{four_color},
\cite{odd_order}, \cite{hales_kepler} which are probably the most well-known
formalisation results in the mathematical community).
We already received numerous invitations to talk about this project in mathematics
departments.
The Notices of the AMS, and their British and French analogues all asked
us to write surveys about our experience as mathematicians using a proof
assistant.
So we can claim that at least some mathematicians noticed our work.

On the other side of the bridge,
we hope our efforts will help encourage the development of
proof assistants to make them easier to use by mathematicians.
Of course a general goal is to have a proof assistant
that combines the expressiveness of dependent types
with the automation level of Isabelle/HOL.
But proving lemmas wasn't actually the most difficult part of our project.
As explained in Section~\ref{sub:Huber_localization},
elaboration can be really difficult,
especially in a context where notations are heavily overloaded.
Lean's current type class system is also put under a lot of pressure
by certain coercions to function types,
and when a given type has multiple structures of the same kind
(like a type being a module over different rings,
or even having different module structures over the same ring).
In addition to those technical issues,
we need more documentation specifically targeting mathematicians.
Such documentation is not easy to write
because of the delicate balance required
between educating mathematicians about computer science
and writing in their native language, but it is crucial.


\vspace{-.1cm}
\paragraph{Acknowledgments}
Although part of the experiment was to see whether a team holding no degree in
logic or computer science could formalize something substantial,
we benefited a lot
from conversations with people in the Lean prover community
who did have such a background,
especially Jeremy Avigad, Mario Carneiro, Johannes Hölzl,
Simon Hudon and Rob Lewis,
and also with Assia Mahboubi and Cyril Cohen,
from the nearby Coq community.
Among mathematicians using Lean,
we received useful comments and
advice from Scott Morrison, Reid Barton and Kenny Lau.
On the informal side of mathematics, we also had useful conversations
with Brian Conrad, and some collaborators of Nicolas Bourbaki who shall
stay anonymous.
We thank Torsten Wedhorn
for making his unpublished notes on adic spaces available --
without this roadmap,
the definition of an adic space would have been much harder to formalise.

\bibliography{perfectoid.bib}
\end{document}